# On the conventional and rotating magnetocaloric effects in multiferroic TbMn$_2$O$_5$ single crystals

M. Balli[1*], S. Jandl[1], P. Fournier[1, 2], D. Z. Dimitrov[3, 4]

[1] Regroupement québécois sur les matériaux de pointe, Département de physique, Université de Sherbrooke, J1K 2R1, QC, Canada.

[2] Canadian Institute for Advanced Research, Toronto, Ontario M5G 1Z8, Canada.

[3] Institute of Solid State Physics, Bulgarian Academy of Science, Sofia 1184, Bulgaria.

[4] Institute of Optical Materials and Technologies, Bulgarian Academy of Sciences, 1113 Sofia, Bulgaria.

**Abstract**.

Solid state-refrigerants have generated worldwide interest owing to their growing potential for use in efficient and green cooling devices. Caloric effects could be obtained by manipulating their degrees of freedom such as magnetization, electric polarization and volume using a variable external field. In conventional magnetocaloric refrigeration systems, the magnetocaloric effect is exploited by moving the active material in and out of the magnetic field source. Here we demonstrate that a giant and reversible magnetocaloric effect can be generated simply by rotating the multiferroic TbMn$_2$O$_5$ single crystal around its b axis in a relatively low constant magnetic field applied in the *ac* plane. For a magnetic field applied along the easy axis *a*, we report an entropy change of 12.25 J/kg K at about 10 K in a field change of 5 T which is 100 times larger than that found when the field is applied along the hard axis *c*. When the TbMn$_2$O$_5$ is rotated with the field remaining in the ac plane, the associated adiabatic temperature change reaches minimum values of 8 K and 14 K under 2 T and 5 T, respectively. This giant rotating magnetocaloric effect in TbMn$_2$O$_5$ is caused by the "colossal" anisotropy of the entropy change, the enhancement of the magnetization under relatively moderate magnetic fields and the lower magnitude of specific heat. On the other hand, the application of the coherent rotational model demonstrates that the rotating magnetocaloric effect exhibited by TbMn$_2$O$_5$ does not originate directly from the magneto-crystalline anisotropy. Our results should inspire and open new ways toward the implementation of compact, efficient and embedded magnetocaloric devices for low temperature and space application. Its potential operating temperature range of 2 to 30K makes it a great candidate for the liquefaction of hydrogen.

# I. Introduction

        Nowadays, the interest in materials with large magnetocaloric effect (MCE) has considerably grown mainly due to their potential use as solid-refrigerants in magnetic cooling systems. Based on the magnetocaloric effect, magnetic refrigeration was recognized to be a serious alternative to traditional refrigerators based on gas compression-expansion owing to its high efficiency as well as its environmental friendliness [1, 2]. The magnetocaloric effect is usually described by the isothermal entropy change ($\Delta S$) and/or the adiabatic temperature change ($\Delta T_{ad}$) when the material is subjected to a varying external magnetic field. It was first implemented by Giauque and MacDougall in the aim to reach very low temperatures achieving successfully temperatures below 1 K by demagnetizing adiabatically a paramagnetic salt [3]. The interest on the magnetic cooling at room temperature has its origin in the reciprocating device reported by Brown in 1976 [4].  By using gadolinium metal (Gd) as the magnetic active substance in an alternating 7 T field produced by an electromagnet, he reached about 46 K no-load temperature span. Aiming to find an alternative for the rare earth Gd and related alloys [5], intensive research has been conducted on magnetocaloric materials leading to the discovery of the giant magnetocaloric effect (GMCE) in $Gd_5Ge_2Si_2$-based compounds associated with a first order magnetic transition (FOMT) [6], which has made a great contribution to the development of magnetic cooling. Few years later, several intermetallic systems undergoing a GMCE near room-temperature, such as MnAs [7, 8], $Fe_2P$-type compounds ($MnFeP_{1-x}As_x$)[9, 10] and $LaFe_{13-x}Si_x$ [11-13] have been successfully found. On the other hand, the manganite-based materials have attracted recently a great attention not only because of their potential use in spintronics applications, but also as magnetic refrigerants due to their large magnetocaloric effect and their high chemical stability [14]. However, the gap to be bridged in going from laboratory samples to a competitive device that meets the market needs is demanding. In fact, the magnetocaloric material must answer a series of requirements before its implementation such as,

sufficiently large MCE, low specific heat, low hysteresis, high electrical resistance, high resistance against oxidation and corrosion, mechanical stability and safe constituent elements [14]. Thus it is very difficult to find a material that combines all these characteristics.

The materials demonstrating interesting magnetocaloric properties in the low temperature range [15-18] are also of great importance from a fundamental point of view and could be used as refrigerants in some specific applications such as space technology and gas liquefaction. According to Barclay et al [16], a broad use of hydrogen for example as fuel and energy carrier will provide better energy security, return major economic, ecological and health benefits. Recently, Matsumoto et al [17, 18], have successfully liquefied hydrogen gas based on the active magnetic refrigeration technique (AMR), in which dysprosium gadolinium aluminium garnet, $Dy_{2.4}Gd_{0.6}Al_5O_{12}$ (DGAG) is used as the magnetic refrigerant. Besides, the manganites $RMnO_3$ and $RMn_2O_5$ (R = rare earth) are actually some of the most promising magnetic refrigerants for working applications in the low temperature range from about 30 K to helium liquid temperatures, due to their ability to satisfy a large part of the requirements cited above [14,19-23]. Additionally, many of these materials as multiferroics exhibit a strong coupling between electric and magnetic order parameters providing new ways for the design of spintronic devices [20-22].

Several studies of $RMn_2O_5$ oxides have demonstrated a large magnetoelectric effect (MEE), which is associated with an unusual commensurate-incommensurate magnetic transition. In the case of $DyMn_2O_5$, the dielectric constant was enhanced by more than 100 % under the application of an external magnetic field leading to a colossal magnetodielectric effect (CMD) [20]. On the other hand, a highly reversible switching of electrical polarization with the help of relatively low magnetic fields was reported in $TbMn_2O_5$ [24]. More recently, a large rotating MCE was reported in the single crystal $HoMn_2O_5$ in the low temperature region, which could open the way for the implementation of compact, simplified and efficient magnetic refrigerators. On the basis of the rotating MCE observed in $HoMn_2O_5$, a new concept for the liquefaction of the helium and hydrogen was also proposed [23].

Usually the multiferroics $RMn_2O_5$ (R = rare earth) are insulators, and display an orthorhombic structure (*Pbam*) composed of $Mn^{4+}O_6$ octahedral and $Mn^{3+}O_5$ pyramidal units [21-22, 24-25]. The octahedra share edges to form chains along the c axis. The formed chains are linked by pairs of pyramids within the ab plane. The interaction between $Mn^{4+}$, $Mn^{3+}$ and rare earth $R^{3+}$ ions magnetic moments lead to a pronounced magnetic frustration resulting in a complex magnetic response [21-22, 24-25]. In this paper, we investigate in details the magnetocaloric properties of $TbMn_2O_5$ single crystals. We demonstrate that in addition to a large conventional MCE, the multiferroic $TbMn_2O_5$ exhibits a giant rotating magnetocaloric effect (RMCE) when it is rotated around the b axis with the relatively low constant applied magnetic field remaining within the *ac* plane. It exceeds largely the RMCE of any known magnetic material, making $TbMn_2O_5$ the best candidate for refrigeration devices working at low temperatures.

## II. Experimental

$TbMn_2O_5$ single crystals were synthesized in a Pt crucible by the high temperature solution growth method using $PbO$-$PbF_2$-$B_2O_3$ flux as described in Ref. 26. Initially, the required polycrystalline samples were prepared from a mixture of $Tb_2O_5$ and $MnO_2$ as the starting materials in stoichiometric quantities, following a conventional solid-state reaction technique. Then, this solid mixture was annealed in open air at 1150 °C for 2 days. The resulting $TbMn_2O_5$ powder was mixed with $PbO$-$PbF_2$-$B_2O_3$ flux and subjected to a heat treatment in a Pt crucible at 1225 °C for 48 h in air. The annealing process was achieved by decreasing the temperature from 1225 °C to 1000°C with the rate of 1°C/h. A Bruker-AXS D8-Discover X-ray diffractometer with the CuKα1 radiation was used to analyse the crystalline structure of the single crystals especially in the θ-2θ and reciprocal mapping modes. The structural characterization was completed by a Raman scattering investigation. Micro-Raman spectra were realized by using a Labram-800 equipped with a microscope, He-Ne laser and a nitrogen-cooled charge coupled device detector (CCD). The magnetization measurements were carried out with the help of a superconducting quantum interference device (SQUID) magnetometer from Quantum Design, model MPMS XL.

## III. Results and discussion

An XRD θ-2θ pattern on a face perpendicular to the c axis of the $TbMn_2O_5$ single crystal is shown in Figure 1 (a). The presence of (00l) peaks confirms the crystal face orientation, its quality with a single domain. Reciprocal space mappings on the (102) and (012) peaks allow us to extract the a and b lattice parameters via their conversion into direct space mappings as illustrated in Fig. 1(b) for the (102) diffraction peak. The analysis of the room temperature XRD pattern shows that the single crystal crystallizes in the orthorhombic structure with space group *Pbam*. The room-temperature lattice constants were found to be a = 7.351 Å, b = 8.602 Å and c = 5.690 Å, which is in agreement with those reported in the literature [26]. Moreover, a detailed examination of the Raman spectra (to be published elsewhere, Mansouri *et al,*) confirms further the orthorhombic structure and the high quality of the $TbMn_2O_5$ single crystals.

The temperature dependence of the zero-field cooled (ZFC) and field-cooled (FC) magnetization curves for $TbMn_2O_5$ in an external magnetic field $\mu_0H = 0.1$ T along the a, b and c axes are shown in Figure 2. The data demonstrate the presence of a large magneto-crystalline anisotropy consistent with that of Ref. 24, where the easy, intermediate and hard magnetization directions are parallel to the a, b and c axes, respectively. This is additionally supported by the isothermal magnetization curves M ($\mu_0H$) measured at 2 K for the main crystallographic directions shown in Fig. 3-a. Similarly, it reveals a gigantic magneto-crystalline anisotropy which is mainly attributed to the large spin-orbit interplay of the Tb moments in $TbMn_2O_5$ single crystals [24]. Also, the magnetic exchange interactions between $Tb^{3+}$, $Mn^{3+}$ and $Mn^{4+}$ moments in $TbMn_2O_5$ cannot all be satisfied simultaneously, leading to a geometrically frustrated magnetic system. The magnetization saturation can be clearly seen for a magnetic field above 2 T applied along the axis a. The corresponding saturation magnetic moment is 140 $Am^2/kg$ (8.75 $\mu_B$/f.u) which is about 52 % of that calculated (268.8 $Am^2/kg$) when considering a parallel configuration of all the magnetic moments of $Tb^{3+}$ (9 $\mu_B$), $Mn^{4+}$ (3.8 $\mu_B$) and $Mn^{3+}$ (4 $\mu_B$). Additionally, the resulting magnetization saturation is very close to the $Tb^{3+}$ magnetic moment (9 $\mu_B$). Moreover, the magnetization at 7T decreases by about 94 % when the magnetic field direction is changed from the easy axis to the hard axis.

It is worth noting that there is practically no significant bifurcation in the temperature dependence of ZFC and FC magnetization curves, indicating the absence of thermal hysteretic effects in $TbMn_2O_5$ singe crystals. With decreasing temperature, a large increase in magnetization can be observed clearly at very low temperatures for the *a* axis, primarily originating from the ferromagnetic ordering of the $Tb^{3+}$ ions [24]. However, as shown in the inset of Fig. 2, the derivative of the magnetization measured as a function of temperature (dM/dT) in a magnetic field of 0.05 T applied parallel to the easy axis *a* reveals anomalies at ~ 5 K and ~ 36 K. The data reported in Fig. 2 combined with the early works [24] indicates that the first anomaly at 5 K corresponds to the ordering of $Tb^{3+}$ moments ($T_{t, Tb}$). The 36 K feature can be ascribed to the appearance of a ferroelectric polarization, consistent also with previous studies [24]. However, additional transitions which occur at 24 K and 40 K can be clearly seen in specific heat measurements [21] but are not visible in the thermomagnetic curves shown in Fig. 2. The 24 K and 40 K transitions have been attributed to the $Mn^{3+}/Mn^{4+}$ spins reorientation and to the long range antiferromagnetic order of the $Mn^{3+}/Mn^{4+}$ spins, respectively. The zero-field-cooled inverse susceptibility ($1/\chi$) as a function of temperature curves, plotted along the main crystallographic direction in a field of 0.1 T are displayed in Fig. 3-b. By fitting the experimental data with the Curie-Weiss law $\chi^{-1}=(T-T_0)/C$ (C is the Curie-Weiss constant) at high temperature shows the Curie-Weiss temperature ($T_0$) being ~ 24K, ~ -162K and -1007 K for the a, b and c axes, respectively. This reflects a "ferromagnetic" ordering along the *a* axis and a strong antiferromagnetic interaction along the *b* and *c* axes. On the other hand, the difference between the Curie-Weiss temperatures for the *a*, *b* and *c* axes confirms the strong magneto-crystalline anisotropy in $TbMn_2O_5$ single crystals. From the linear fit of $\chi^{-1}(T)$, we have also deduced the effective magnetic moment of 11.51 $\mu_B$/f.u. along the a axis, which is in perfect agreement with the theoretically expected value:

$$\mu_{eff} = \sqrt{(\mu_{eff}(Tb^{3+}))^2 + (\mu_{eff}(Mn^{3+}))^2 + (\mu_{eff}(Mn^{4+}))^2} = 11.59 \mu_B.$$

To understand further the nature of the phase transition associated with the magnetic moment ordering of $Tb^{3+}$, we have used the criterion given by Banerjee [27] to distinguish second-order magnetic transitions from first-order ones. This is important since the magnetocaloric properties can be strongly affected by the nature of the magnetic transition. The Banerjee approach consists in the investigation of the slope of H/*M* versus $M^2$ isotherm curves, also known as Arrott plots [28]. For the second order transition, the slope of H/*M* versus $M^2$ remains positive for all temperatures, whereas it is negative for a FOMT. Our analysis of magnetization isotherms for $TbMn_2O_5$ single crystals along the easy axis (above and below $T_{t, Tb}$) (not shown here) confirms the absence of any negative slope at low as well as high external magnetic field revealing that the phase transition corresponding to $Tb^{3+}$ magnetic moments ordering is second-order in nature.

The main purpose of this paper is to investigate the magnetocaloric properties of TbMn$_2$O$_5$ with a great attention paid to its rotating MCE as was done previously for HoMn$_2$O$_5$ [23]. In addition to the adiabatic temperature change ($\Delta T_{ad}$), the temperature dependence of isothermal entropy change ($\Delta S$) due to the variation of applied magnetic field is the most used parameter for evaluating the magnetocaloric potential of a magnetic material. In an isothermal process of magnetization, $\Delta S$ can be obtained from isotherms curves (M vs H) by using the well-known Maxwell relation,

$$\Delta S(T, 0 \rightarrow H) = \int_0^H \left( \frac{\partial M}{\partial T} \right)_{H'} dH' \qquad (1)$$

Due to the discreetness of the data, the $\Delta S$ values are usually computed by the following numerical formula,

$$\Delta S = \sum_i \frac{M_{i+1} - M_i}{T_{i+1} - T_i} \Delta H_i \qquad (2)$$

where $M_{i+1}$ and $M_i$ are the measured magnetizations in a field $H$, at temperatures $T_{i+1}$ and $T_i$, respectively. The isothermal M (H) curves of the single crystal TbMn$_2$O$_5$ measured as a function of magnetic field applied along the easy, intermediate and hard axes at different fixed temperatures are plotted in Figure 4. Along the easy axis $a$, the magnetization isotherms demonstrate a typical "ferromagnetic" behaviour at temperature below T$_{t,Tb}$, while the linear field dependence of the magnetization at high temperature is typical for a "paramagnetic phase". Magnetization measurements were also carried out in increasing and decreasing magnetic field close to the transition temperature T$_{t,Tb}$. The obtained loops (not shown here) does not show any hysteresis effect in any of the field direction revealing the perfect reversibility of the MCE in TbMn$_2$O$_5$ which is a favourable situation from a practical point of view. It is worth noting that the applicability of the Maxwell relation for calculating MCE has been controversial in recent years [29- 32], especially for materials presenting a first order magnetic transition associated with a large hysteresis. This is obviously not the case for the TbMn$_2$O$_5$ single crystal under investigation in the present study.

Figures 5-a to c show the computed value of –$\Delta S$ as a function of temperature under several external magnetic fields along the $a$, $b$ and $c$ axes. As shown, the isothermal entropy change

reveals also a considerable anisotropy. As shown in Fig. 6-a, the maximum value of $-\Delta S$ along the easy axis in a field of 7 T is about 63 times larger than that along the hard orientation. For H//a, the $-\Delta S$ maxima are distributed in the temperature range between $\sim$ 5K and $\sim$ 15K. When changing the magnetic field from 0 to 2 T, 0 to 5 T and 0 to 7 T, the isothermal entropy change peak reaches values of 6.4, 11.4 and 13.35 J/kg K, for H//a, while it is only 0.33 J/kg K, 1.8 J/kg K and 3 J/kg K for H//b, respectively. For the c axis, the corresponding entropy change is negligible, even for high magnetic fields ($\sim$ 0.2 J/kg K for 7 T). Considering the case in which the disordered magnetic phase of $Tb^{3+}$ ions is changed to a completely ordered phase, the theoretical limit of the resulting entropy change is given by $\Delta S_{Limt}=R*Ln(2J+1) = 61. 18$ J/kg K (here, $R$ is the universal gas constant and J is the angular momentum quantum number). As for $Tb^{3+}$, J was assumed to be 6 for the $TbMn_2O_5$ single crystal. For a magnetic field variation from 0 to 7 T along the ordering axis of $Tb^{3+}$ moments (axis a), only 22 % of $\Delta S_{Limt}$ can be obtained upon saturation of the magnetization, revealing that entropy changes larger than 13.35 J/kg K may be reached under intense external magnetic fields.

For the implementation of the magnetocaloric effect in cooling devices, it is suitable to search for magnetic materials that not only show large MCE with low thermal and field hysteresis losses, but also have a large refrigerant capacity (RC). This parameter has been suggested by Gschneidner and Pecharsky [33] as an important figure of merit for the evaluation of magnetocaloric materials. RC which is a measure of the amount of thermal energy that can be transferred between the hot ($T_H$) and cold ($T_C$) reservoirs in one ideal refrigeration cycle is defined as

$$RC = \int_{T_C}^{T_H} \Delta S(T)dT \,. \qquad (3)$$

$T_H$ and $T_C$ are the temperatures defining the width at half maximum of the $-\Delta S$ (T) profile. The RC values as a function of applied magnetic field along the main crystallographic directions is reported in Fig.6-b. The considerable difference between the RC values along the principal crystallographic axes is another manifestation of the huge anisotropy. With increasing field, the refrigerant capacity increases almost linearly with a rate of about 79 J/kg T, 8 J/kg T for H//a and H//b, respectively. Along the hard direction, the RC is negligible. Under a magnetic field change of 0-7 T, the RC values for $TbMn_2O_5$ is 480 J/kg, 47 J/kg and only about 4 J/kg along the a, b and c axes, respectively. Considering magnetocaloric oxides with similar working temperature range, the magnetocaloric properties of $TbMnO_3$ were recently investigated and reported in Ref. 33. Although the $TbMnO_3$ and $TbMn_2O_5$ single crystals contain the same rare earth element ($Tb^{3+}$) and present similar magnetization saturation (about 140 $Am^2$/kg), it was found that the maximum isothermal entropy change under 7 T along the easy axis a  for $TbMnO_3$ is larger (18 J/kg K) than that exhibited by $TbMn_2O_5$ (13.35 J/kg K). However, the refrigerant capacity for $TbMn_2O_5$ is much larger than that for $TbMnO_3$ (390 J/kg for 7 T //a). Theses observed deviations originate mainly from the significant difference in their magnetic structure. In fact, the antiferromagnetic ground state is not quite stable in $TbMnO_3$ against applied field along the easy axis *a* [34]. At low temperatures, the spiral magnetic structure of $Tb^{3+}$ and $Mn^{3+}$ ions in $TbMnO_3$ enables to induce easily a first-order magnetic transition of $Tb^{3+}$ moments from the antiferromagnetic state to the ferromagnetic phase. Such metamagnetic-like transformation means that the magnetization can be rapidly changed with varying magnetic field leading to a large magnetic entropy change. In contrast with $TbMnO_3$, the magnetic phase transition associated with the magnetic moments ordering of $Tb^{3+}$ in the single crystal $TbMn_2O_5$ shows a second-order character. This makes the magnetic transition broader, leading to a large refrigerant capacity in $TbMn_2O_5$ single crystals. On the other hand, the RC of the $TbMn_2O_5$ single crystal  (314 J/kg for 5 T) along the easy axis is much larger than that found in the oxides $HoMn_2O_5$ (334 J/kg for 7 T) [23], $DyMnO_3$ (225 J/kg for 7 T) [35] and $HoMnO_3$ (120 J/kg for 5 T) [36], and even in some intermetallics such as  $ErRu_2Si_2$ (196.5 J/kg for 5 T) [36], $DySb$ (144 J/kg for 5 T) [38] and $ErMn_2Si_2$ (273 J/kg for 5 T) [39]. The reversible "conventional" magnetocaloric effect combined with a large refrigerant capacity make $TbMn_2O_5$ a serious candidate for magnetocaloric-based devices operating at low temperatures.

As demonstrated above, the single crystal $TbMn_2O_5$ reveals a large "conventional" magnetocaloric effect which can be obtained by exposing the material to a varying external magnetic field. However, as shown in Fig. 6-a, the MCE represented by the isothermal entropy change exhibits a "colossal" anisotropy in $TbMn_2O_5$ single crystals.  This means that a large MCE can be also generated by spinning$TbMn_2O_5$ crystals around their b axis in a constant magnetic field maintained within its *ac* plane (Fig.7-a), thus without the need to vary the magnitude of the applied magnetic field. It is worth noting that the studies of the rotating (or anisotropic) MCE are relatively rare [23, 34, 40-42]. In a previous work, we have proposed a new design for the liquefaction of helium and hydrogen based on the large anisotropy of the MCE in $HoMn_2O_5$ single crystals [23]. Besides, magnetocaloric materials are studied mainly around the ordered-disordered transition point under magnetic field variations. This was attributed to the fact that the contribution of the magnetocrystalline anisotropy to MCE at the magnetic phase transition is much lower than that generated by the change in magnetic order [41]. For different reasons, the implementation of the rotating MCE (RMCE) could revolutionize research and development on magnetic cooling technology particularly for low temperature and space applications: 1) In magnetic cooling systems using standard MCE, the magnetization-demagnetization process when using standard magnetocaloric effect requires generally a large mechanical energy for moving the active material in and out of the magnetic field zone, decreasing consequently the system efficiency. Hence, the use of the RMCE would enable the reduction of the energy absorbed by the cooling machine. 2) The implementation of such effect allows the conception of rotary magnetic refrigerators working at high frequency leading to a large cooling power [43]. 3) The continuous variation of the magnetic field in cooling systems leads to the appearance of electric currents in metallic refrigerant materials. RMCE in a constant magnetic field eliminates the energy losses and additional works caused by the resulting eddy currents [44]. 4) It is known that rotary magnetic refrigerators are more efficient than reciprocating devices [43]. However, for rotary systems using the "standard MCE", the need to create a magnetic field gradient makes the design of the magnetic field source and consequently the cooling machine more complex [43]. Therefore, the design of the machine can be drastically simplified by the implementation of materials exhibiting a large anisotropic MCE, since this kind of devices requires a simple constant magnetic field source that would lead to more compact setups [23]. 5) The implementation of the RMCE can also be of benefit from an economical point of view, since the rotating motion can be easily realized with the help of cheaper circular motors.

In this perspective, we have first estimated the entropy change corresponding to the rotation of the single crystal TbMn$_2$O$_5$ from the hard axis $c$ to the easy axis $a$ ($\Delta S_{R,ca}$) in a constant magnetic field. By setting initially the magnetic field along the c axis as the starting direction, the rotating entropy change when the single crystal is rotated by an angle of 90° in the $ac$ plane under a constant magnetic field H can be expressed as follows [23, 34, 40] : $\Delta S_{R,ca} = \Delta S$ (H//a)- $\Delta S$ (H//c), where $\Delta S$ (H//a) and $\Delta S$ (H//c) are the entropy changes when the magnetic field is applied along the a and c axes, respectively. The temperature dependence of $\Delta S_{R,ca}$ under several constant magnetic fields for TbMn$_2$O$_5$ is reported in Figure 7-b. At 7 T, the maximum value of $\Delta S_{R,ca}$ is evaluated to be 13.35 J/kg K, which is larger than the rotating entropy change reported in TmMnO$_3$ (8.73 J/kg K) [40] and TbMnO$_3$ (8.2 J/kg K) [34] single crystals in similar working temperature spans. Additionally, the difference of RC between the easy and hard axes reaches about 476 J/kg (7 T) for TbMn$_2$O$_5$, while it is only 304.1 J/kg (7 T) in the case of TbMnO$_3$ [33] and 205.4 J/kg (7 T) for TmMnO$_3$ [40].

According to the recently reported rotating entropy change in TmMnO$_3$, TbMnO$_3$ and HoMn$_2$O$_5$ single crystals [23, 34, 40], one can clearly see that high magnetic fields must be used to generate enough RMCE. However, one of the important advances concerning the RMCE and reported in this paper is that TbMn$_2$O$_5$ reveals a large RMCE at relatively low magnetic fields ($\mu_0$H ≤ 3 T). At 2 T, the maximum value of $\Delta S_R$ exhibited by TbMn$_2$O$_5$ is 6.4 J/kg K, which is about 3 times larger than that found in TbMnO$_3$ [34] and 6 times higher than that exhibited by TmMnO$_3$ [40]. The behavior of the RMCE as a function of the magnetic field for both TbMn$_2$O$_5$ and HoMn$_2$O$_5$ are also compared in Fig.7-c. As shown, the rotating entropy change is enhanced in TbMn$_2$O$_5$, particularly for low magnetic fields (more than 2 times larger) although the magnetic moment of Tb$^{3+}$ (9 $\mu_B$) is smaller than that of Ho$^{3+}$ (10 $\mu_B$) [45]. The presence of a large MCE under low fields in TbMn$_2$O$_5$ is of great importance from a practical point of view since the needed magnetic field in MCE-based devices can be provided simply by permanent magnets with an upper limit value of around 2 T [46-48]. The enhancement of the rotating entropy change in TbMn$_2$O$_5$ compared with the previous reported materials, can be attributed mainly to the gigantic magneto-crystalline anisotropy in TbMn$_2$O$_5$ resulting from the large spin-orbit interplay of Tb moments. Also, the possibility to saturate the magnetization in TbMn$_2$O$_5$ along the easy direction $a$ with relatively low magnetic fields (around 2 T), contributes for this enhancement.

It is known that the entropy change $\Delta S$ is the magnetocaloric parameter which determines the capacity of a material to absorb or generate thermal energy. However, the adiabatic temperature change $\Delta T_{ad}$ is also of great importance for the evaluation of magnetocaloric materials, since it defines the maximum temperature span which that a magnetic material can achieve in a magnetocaloric device. The rotating adiabatic temperature change can be roughly estimated as follow [23, 34, 49]:

$$\Delta T_{R,ad} = -\frac{T}{C_P(H=0)}\Delta S_R \qquad (4)$$

where C$_p$ is the specific heat. The rotating (from c to a) adiabatic temperature change can be also obtained by using the equation:

$$\Delta T_{R,ad}(T,H) = [T(S)_{H//a} - T(S)_{H//c}]_S \qquad (5)$$

For this purpose, the specific heat measurements reported in Ref.21 at zero-magnetic field were used. The full entropy at zero-field can be obtained from specific heat data using the equation:

$$S(0,T) = \int_0^T \frac{C_P(0,T^{'})}{T^{'}} dT^{'} \qquad (6)$$

S (H, T) curves can then be constructed by subtracting the related isothermal entropy change ΔS (H, T) obtained using the magnetization measurements from the S (0, T) curve as given in Figure 8-a for a magnetic field of 5T applied along the a and c axes.

The resulting $\Delta T_{ad}$ as a function of temperature when the single crystal $TbMn_2O_5$ is rotated by 90 ° from the a to the c axis is reported in Fig. 8-b for 2 and 5T. From Fig.8-b, the maximum $\Delta T_{ad,ca}$ reaches values of 8 and 14 K for a magnetic field of 2 and 5 T, respectively. On the basis of Eq.4, the maximum $\Delta T_{ad,ca}$ was found to be about 9.5 K and 18.7 K for 2 and 5 T, respectively (Fig-9). Considering magnetocaloric materials with a similar working temperature range, $TbMn_2O_5$ exhibits a rotating adiabatic temperature change at least 3 times larger than that exhibited by $TbMnO_3$ [34] and $HoMn_2O_5$ [23] at 5 T. Considering low magnetic fields, $\Delta T_{ad,ca}$ presented by $TbMn_2O_5$ is more than 5 times higher than that exhibited by $TbMnO_3$ [34] and $HoMn_2O_5$ [23] single crystals at 2 T. The giant $\Delta T_{ad,\,ca}$ found in $TbMn_2O_5$ can be considered as a major advance in the field of materials with RMCE. On the other hand, since the entropy change along the hard axis is negligible [S (0 T) = S (H//c)] the adiabatic temperature change resulting from the magnetization of $TbMn_2O_5$ along the easy axis is similar to the $\Delta T_{ad,ca}$ reported in Fig. 8-b. The enhancement of the rotating adiabatic temperature change in the single crystal $TbMn_2O_5$ can be attributed once again to the "colossal" anisotropy of the entropy change in $TbMn_2O_5$, the possibility to saturate the magnetization at relatively low magnetic field and particularly the low value of the specific heat. Near the ordering temperature of the rare earth ($Tb^{3+}$, $Ho^{3+}$), the reported specific heat is about 15 J/kg K for $TbMnO_3$ [34, 50] and 19 J/kg K for $HoMn_2O_5$ [21] while it is only about 6.5 J/Kg K for $TbMn_2O_5$ [21]. According to Eq.4, the lower the specific heat is, the higher the adiabatic temperature change may be.

Considering now the magnetic field initially parallel to the easy axis *a*, a negative RMCE can be obtained by rotating the single crystal $TbMn_2O_5$ from the *a* to the *c* direction by 90 °. The associated adiabatic temperature changes curves ($\Delta T_{ad,ac}$) in different constant magnetic fields can be generated from Fig. 8-a or created with the data of $\Delta T_{ad,ca}$ shown in Fig 8-b. In fact, if the initial temperature of the magnetocaloric material is $T_I$ in the case when the magnetic field is parallel to the axis c, then the temperature increases by $\Delta T_{ad,ca}$ (in adiabatic conditions) due to the rotation from the c axis to the a axis (90°), leading to a final temperature of $T_F = T_I + \Delta T_{ad,\,ca}$. Therefore, at $T_F$ the rotation of TbMn$_2$O$_5$ in the inverse direction will decrease the material temperature from $T_F$ to $T_I$ by $\Delta T_{ad,ac}$, which results in a negative adiabatic temperature change, $\Delta T_{ad,ac}$. The derived $\Delta T_{ad,ac}$ as a function of temperature is also shown in Fig. 8-b for a magnetic field of 5 T (negative values). Around 20 K, hydrogen's liquid temperature, a simple rotation of TbMn$_2$O$_5$ in 5 T enables to decrease its temperature from about 20 K to about 5 K. This is of great importance from a practical point of view, since it may allow the design of efficient magnetic liquefiers. It is worth noting that Matsumoto et al. [17] and Numazawa et al. [18] have reported more recently a magnetic refrigerator for hydrogen liquefaction. A ceramic magnetic material, dysprosium gadolinium aluminium garnet Dy$_{2.4}$Gd$_{0.6}$Al$_5$O$_{12}$ (DGAG) was utilized as a refrigerant. The large specific heat of this kind of materials limits the adiabatic temperature change to typically $1 \sim 2$ K/T [17]. On the other hand, the active magnetic refrigeration bed constituted of DGAG is moved inside and outside of a superconducting magnet. Furthermore, the use of TbMn$_2$O$_5$ single crystals which exhibit a large adiabatic temperature change ( more than 8 K for 2 T) that can be obtained through a rotation process (instead of magnetization-demagnetization), could make (for example) the hydrogen liquefier reported in Refs.17-18 more efficient and compact with a simplified design.

In order to understand the physics behind the rotating magnetocaloric effect in TbMn$_2$O$_5$, we have used the coherent rotational model to determine the contribution of the magneto-crystalline anisotropy to the rotating entropy change [23, 34, 40, 45]. Assuming the uniaxial character of the magnetocrystalline anisotropy, the energy of the single crystal can be expressed as follows:

$$E = E_K + E_H = K_1 \sin^2 \alpha + K_2 \sin^4 \alpha - \mu_0 M_0 H \cos(\beta - \alpha) \qquad (7)$$

$K_1$ and $K_2$ represent the anisotropy constants of second and fourth orders, $M_0$ the magnetization saturation, α is the angle between the easy axis $a$ and the magnetization orientation and finally, β is the angle between the magnetic field and the easy axis. $K_1$ and $K_2$ can be estimated from the equation:

$$\mu_0 H = \frac{2K_1}{M_0^{\,2}} M + \frac{4K_2}{M_0^{\,4}} M^3 \qquad (8)$$

which is obtained by minimizing Eq. (7) for β = 90°. Owing to the linear relationship between M and $\mu_0$H along the hard direction (Fig.4-b) $K_2$ can be assumed equal to zero and $K_1$ can be found by fitting Eq. (8)

to magnetization isotherms reported in Fig. 4-b. Considering the magnetic field initially parallel to the hard axis ($\beta = 90°$), the isothermal rotating entropy change resulting from the magneto-crystalline anisotropy when $TbMn_2O_5$ is rotated from the angle 90 ° to $\beta$ can be written as [23, 34, 40, 45]:

$$\Delta S_{MCA}(H,\beta) = \int_{90°}^{\beta} \frac{dE_K}{T} = K_1 \frac{\sin^2 \alpha(H,\beta) - \sin^2 \alpha(H,\beta = 90°)}{T} \qquad (9)$$

According to our magnetic measurements, the magnetization saturation is about 140 Am$^2$/kg (Fig.2(b)) and the anisotropy constant $K_1$ was found to be 8032 J/kg at 11 K. The values of the angle $\alpha$ can be obtained as a function of the magnetic field and $\beta$ at equilibrium state by minimizing Eq. (7). $\Delta S_{MCA}$ is only 2.87 J/kg K when the crystal is rotated from 90 ° to $\beta = 0$ ° under 7 T, which is much smaller than the experimental value determined from Maxwell's relation (about 13.35 J/kg K). This discrepancy can be attributed to the complexity of the magneto-crystalline anisotropy in the frustrated magnetic material $TbMn_2O_5$. On the other hand, other contributions to the rotating magnetocaloric effect such as thermal fluctuations must be considered. In order to answer this question, magnetization measurements as a function of the angle $\beta$ must be performed, which will be the subject of a future study. However, by fitting the Eq. 9 with experimental data reported in Fig. 7-b, an "effective" value of "1680 J/kg" was determined for the anisotropy constant $K_1$ for T = 11 K and $\mu_0 H = 7$ T. Using this effective value, the expected behaviour for the rotating entropy change as a function of $\beta$ (under 7 T) is reproduced in Fig. 10 for a complete rotation (rotation by 360 °) of the single crystal $TbMn_2O_5$. Such data are of great importance when modelling the performance of an AMR thermodynamic cycle in a functional device. As shown in Fig. 10, a magnetic liquefier can be built by rotating continuously (by 90°) an array of single crystals of $TbMn_2O_5$ around their c axis while keeping the constant magnetic field in the *ac* plane. The cooling process can be divided into four steps:

1) When the crystal is rotated from $\beta = 90°$ to $\beta = 0°$, it undergoes a reduction in its magnetic entropy, increasing consequently the material's temperature.

2) During that part of the cycle, the generated heat can be evacuated outside the system using a secondary heat transfer fluid (gas).

3) As the crystal is further rotated from $\beta = 0°$ towards $\beta = - 90°$, the crystal cools down due to the increase of its magnetic entropy.

4) During that part of the cycle, the fluid to be liquefied (He, H) flows through the blocks of $TbMn_2O_5$ in order to extract the heat out of the fluid.

This thermodynamic cycle can be repeated at high frequency without the losses generally associated with eddy currents (which are obviously dependent on frequency), making a very efficient process.

## IV. Conclusion

In summary, we examined the anisotropy-induced rotating magnetocaloric effect in magnetoelectric $TbMn_2O_5$ single crystals. The "colossal" anisotropy of the entropy change, the magnetization enhancement, the absence of hysteresis and the low specific heat leads to a giant and reversible rotating magnetocaloric effect in $TbMn_2O_5$ at a relatively moderate magnetic field of 2T, reachable with permanent magnets. Entropy changes of 6.35 J/kg K and 12.25 J/kg K and adiabatic temperature changes larger than 8 K and 14 K can be reached simply by rotating $TbMn_2O_5$ around its b axis in constant magnetic fields of 2 T and 5 T applied in the *ca* plane, respectively. This means that the refrigeration process can be achieved by rotating continuously $TbMn_2O_5$ crystals in a magnetic field, which is preferred than moving it in and out of the magnetic field region. Furthermore, following the coherent rotational model, we determined that the RMCE in $TbMn_2O_5$ is not uniquely caused by the magneto-crystalline anisotropy. The obtained results combined with the insulating character of $TbMn_2O_5$ open ways for the development of a new generation of magnetic cooling devices especially for low temperature and space applications. However, the concept of RMCE-based devices could be also used for room temperature tasks. The challenge now is seeking for materials with similarly large RMCE at ambient temperature.

**Acknowledgments**

The authors thank M. Castonguay and S. Pelletier for technical support and S Mansouri for the Raman spectra of TbMn$_2$O$_5$. We acknowledge the financial support from NSERC (Canada), FQRNT (Québec), CFI , CIFAR and the Université de Sherbrooke.

**Figure captions**

**Figure 1:** XRD patterns for the single crystal TbMn$_2$O$_5$ at room temperature: (a) θ-2θ pattern with the *c* axis normal to the face of the crystal; (b) Reciprocal space mapping converted into a direct space mapping of the (102) peak of the same crystal on the same *c*-axis face giving directly the lattice parameters.

**Figure 2:** Temperature dependence of ZFC and FC magnetization of TbMn$_2$O$_5$ under a magnetic field of 0.1 T along the main crystallographic axes. Inset: derivative of the magnetization measured as a function of temperature in a magnetic field of 0.05 T along the easy axis.

**Figure 3:** (a) Isothermal magnetization curves measured at 2 K for H//*a*, H//*b* and H//*c*. (b) 0.1T-ZFC inverse susceptibility as a function of temperature.

**Figure 4:** Isothermal magnetization curves of the single crystal TbMn$_2$O$_5$ for (a) H//*a,* (b) H//*b* and (b) H//*c*. The increments of temperature are 2 K from 2 to 26 K and 4 K from 26-82 K.

**Figure 5:** Temperature dependence of the isothermal entropy change in TbMn$_2$O$_5$ under different magnetic fields for (a) H//*a*, (b) H//*b* and (c) H//*c*.

**Figure 6:** (a) Comparison between entropy changes related to H//*a*, H//*b* and H//*c* for 7 T. (b) Refrigerant capacity corresponding to the three main axes as a function of the magnetic field.

**Figure 7:** (a) Generation of the MCE by rotating TbMn$_2$O$_5$ single crystals in the ac plane. (b) Entropy changes related to the rotation of the single crystal TbMn$_2$O$_5$ from the *c* to the *a* direction by 90 ° in different constant magnetic fields, with magnetic field initially parallel to the *c* axis. (c) Magnetic field dependence of the maximum rotating entropy change in TbMn$_2$O$_5$ (triangles) and HoMn$_2$O$_5$ (open circles, data taken from Ref.22).

**Figure 8:** (a) Temperature dependence of the full entropy in a magnetic field of 5 T applied along the hard and easy axes. (b) Deduced rotating adiabatic temperature change as a function of temperature.

**Figure 9:** Rotating adiabatic temperature change as a function of temperature, deduced from Eq.4.

**Figure 10:** Rotating entropy change calculated from the coherent rotational model and corresponding to a complete rotation (rotation by 360 °) of TbMn$_2$O$_5$ in the *ca* plane.

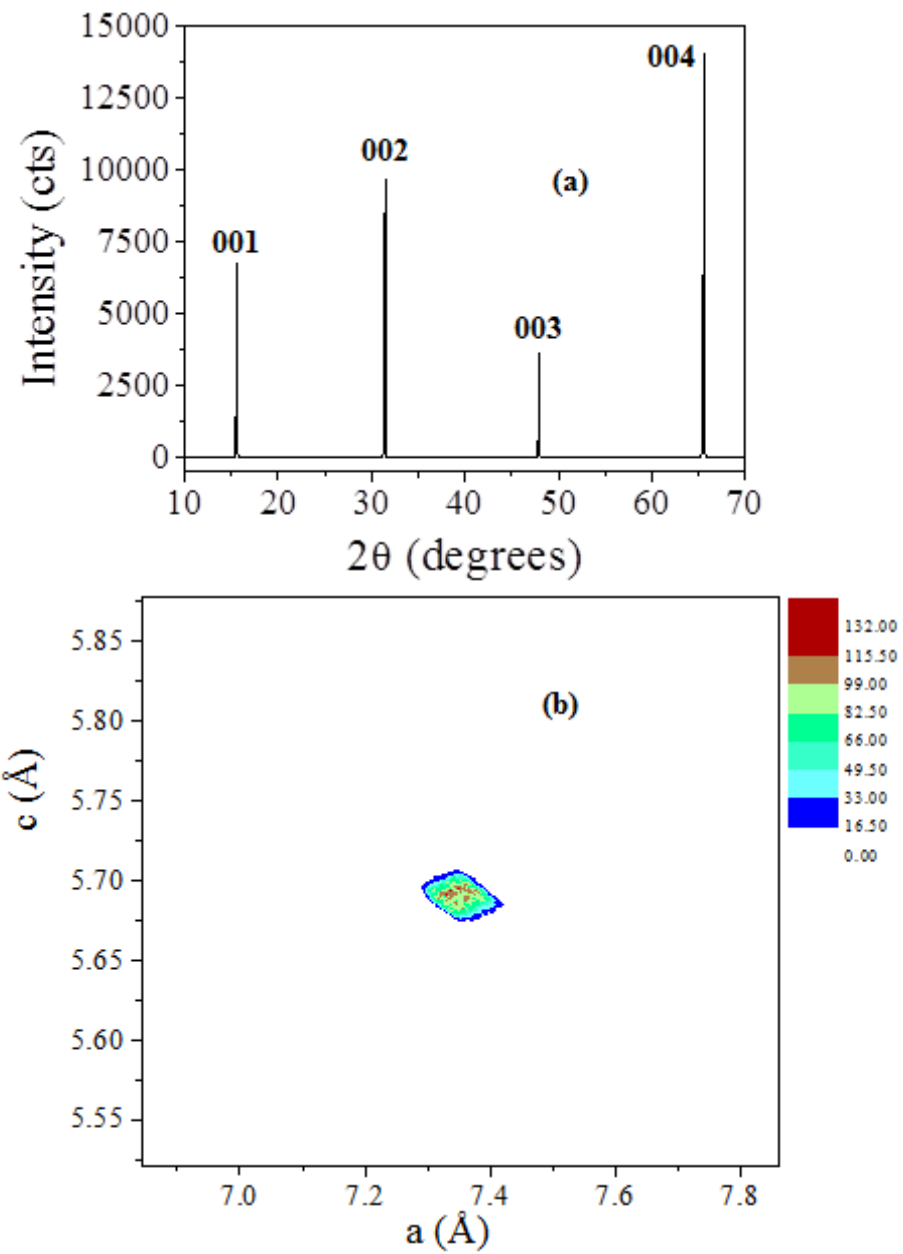

**Figure 1**

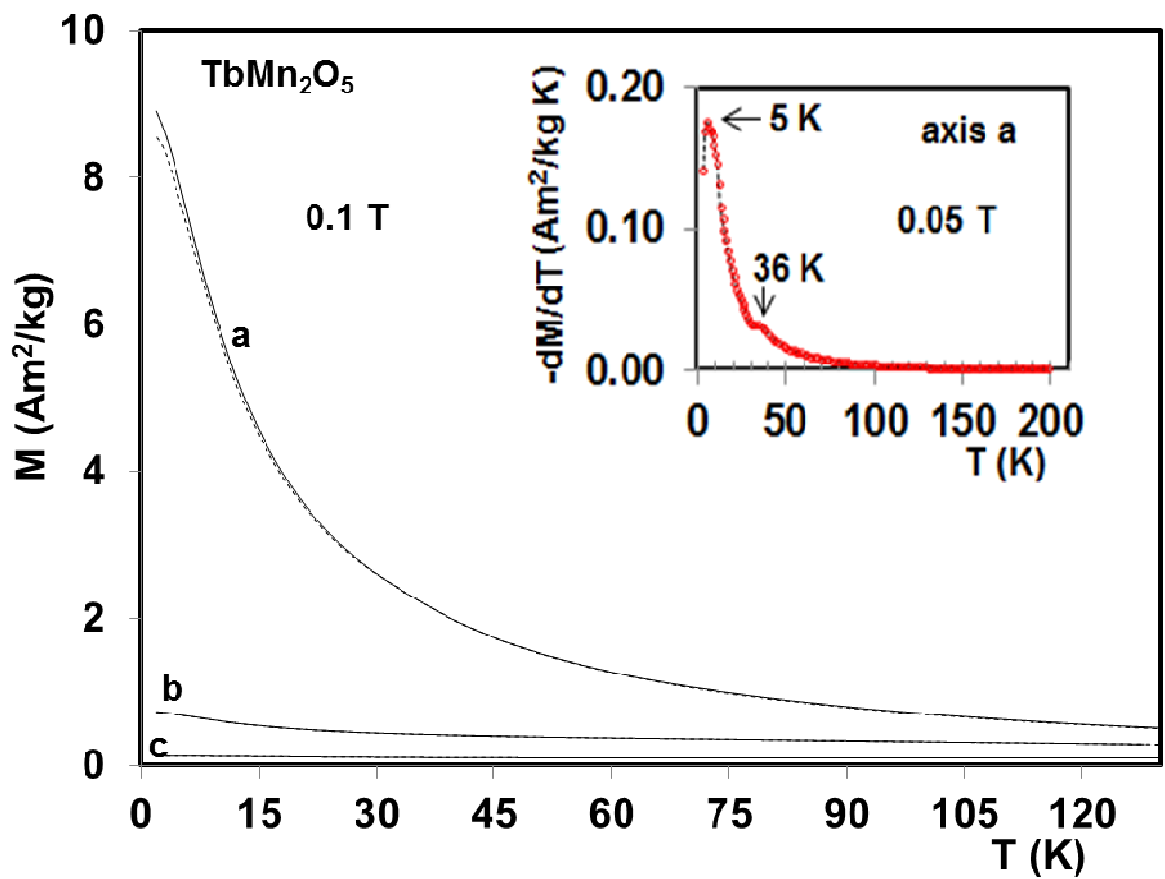

Figure 2

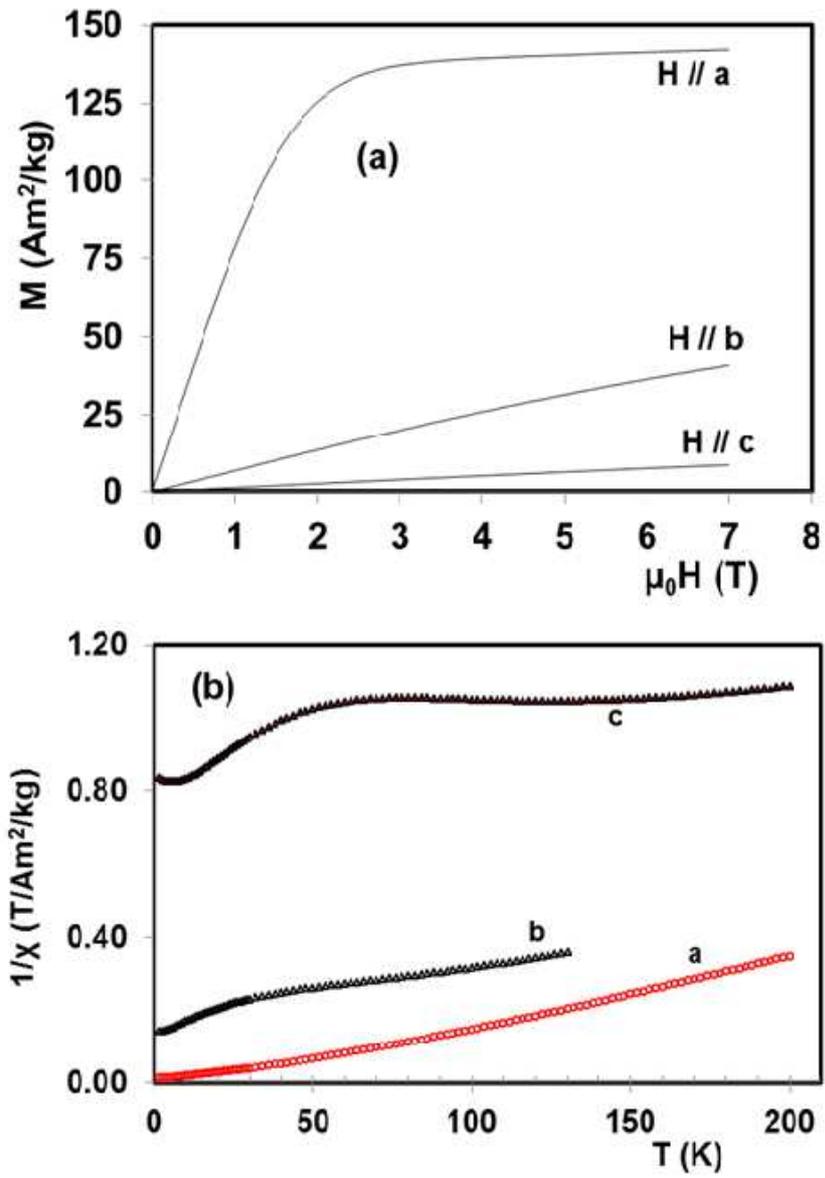

**Figure 3**

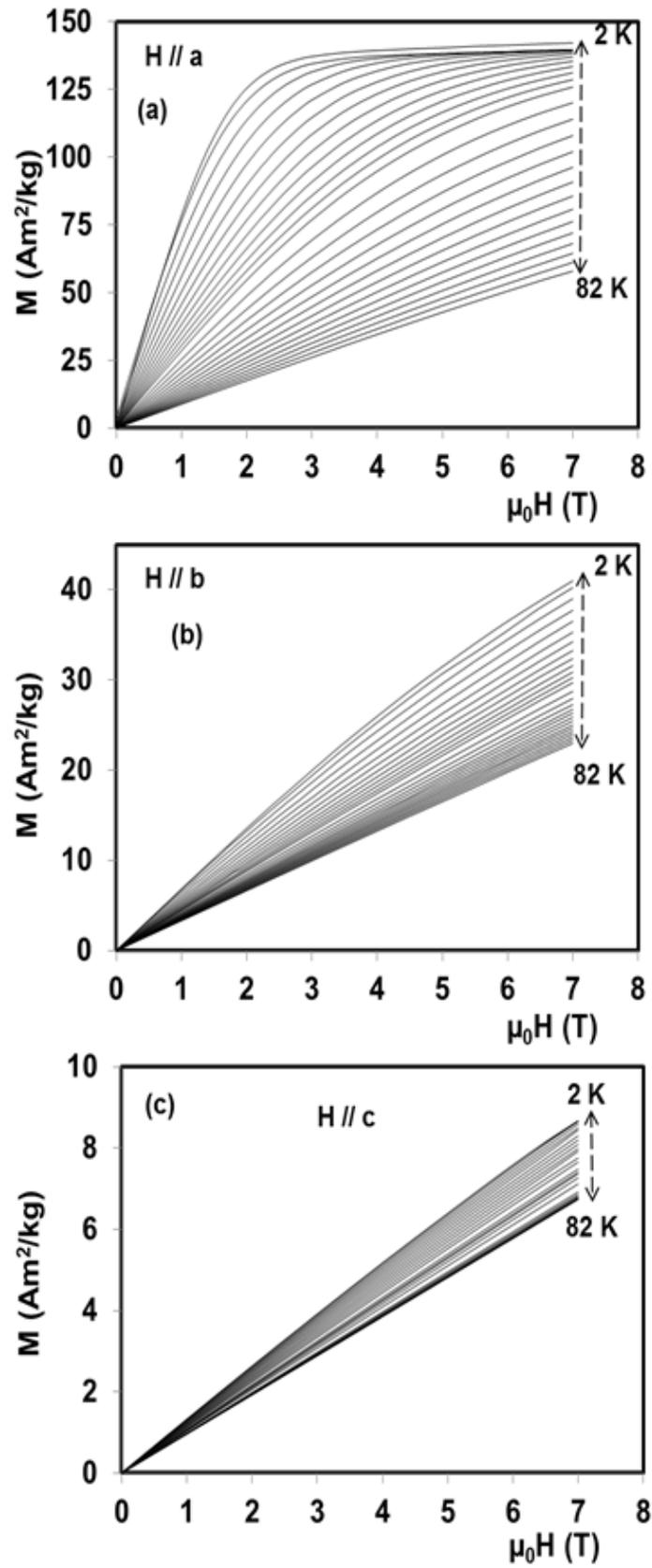

**Figure 4**

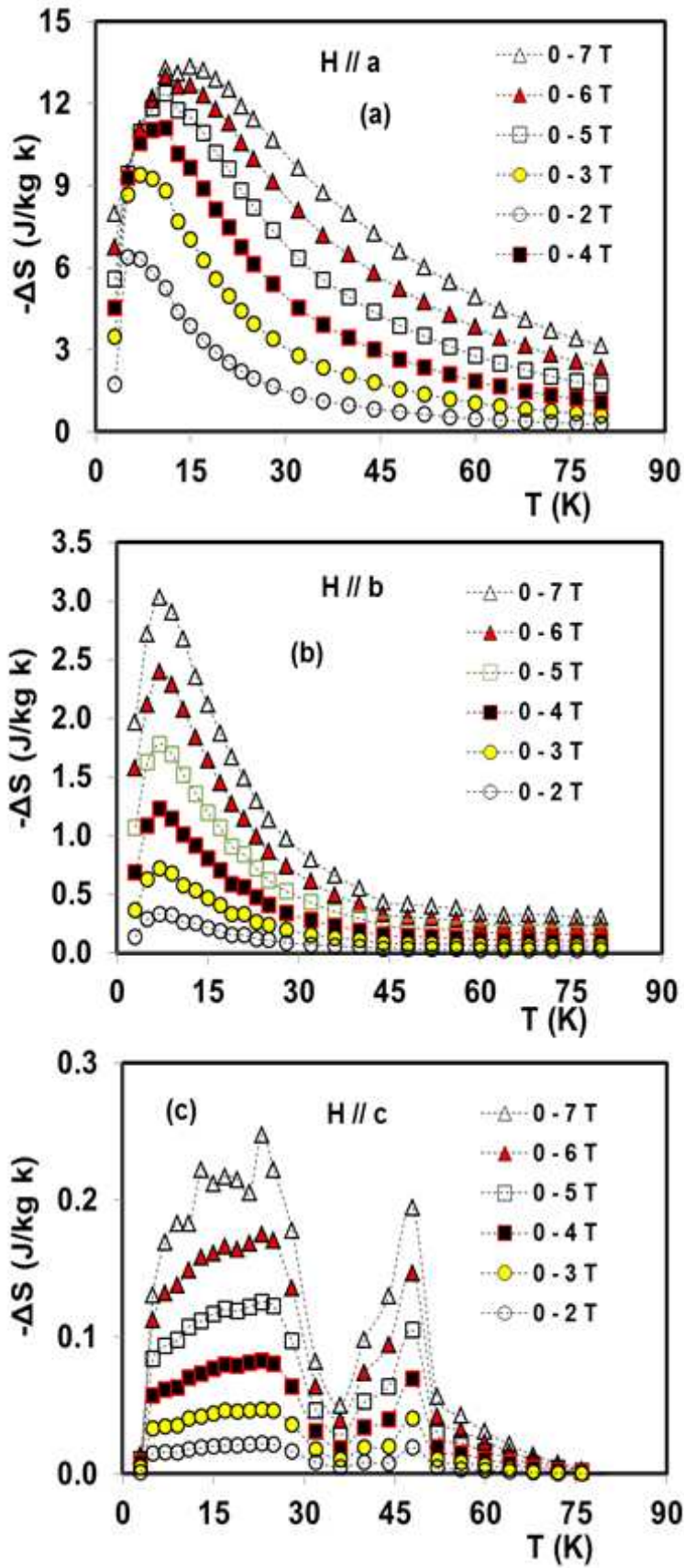

**Figure 5**

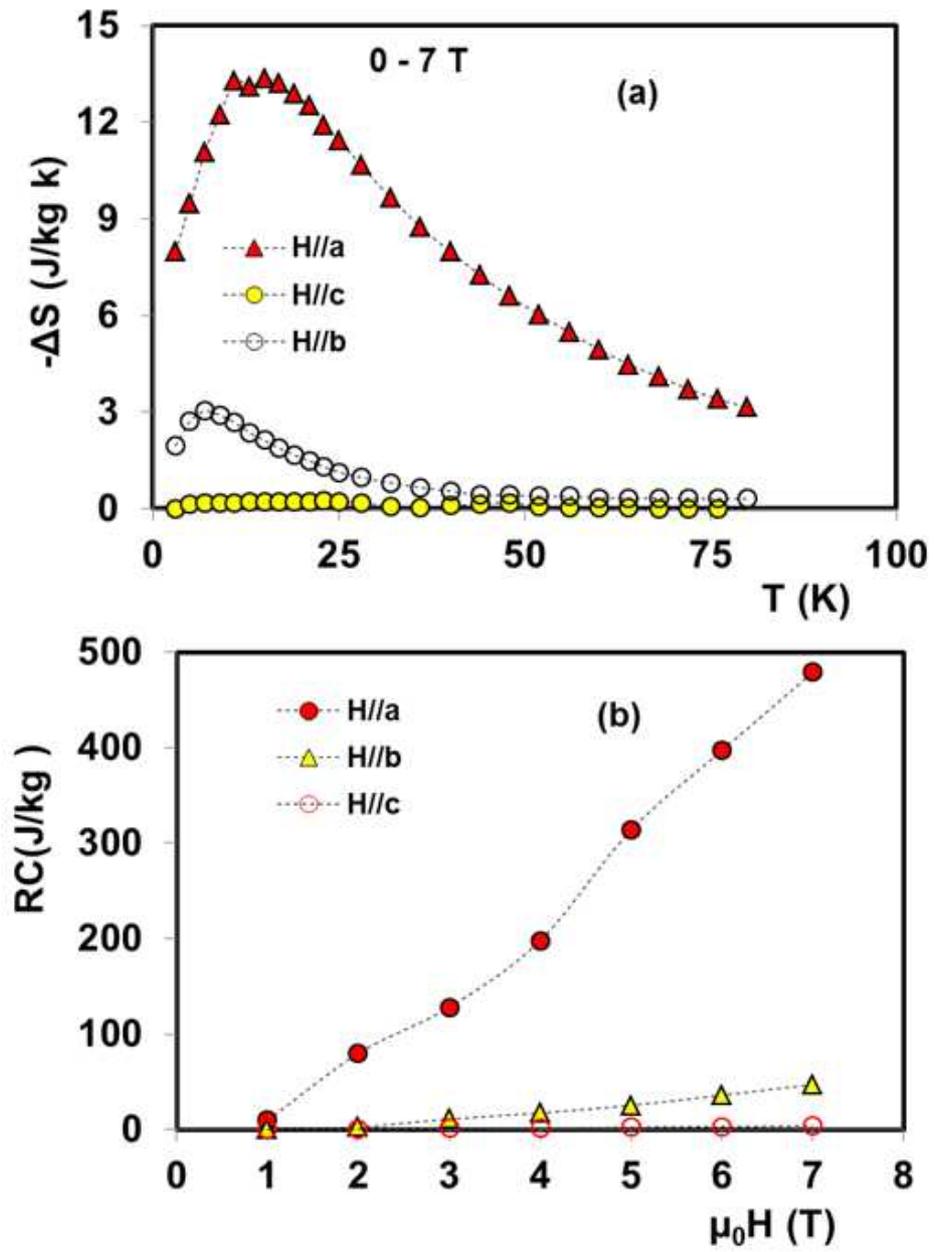

Figure 6

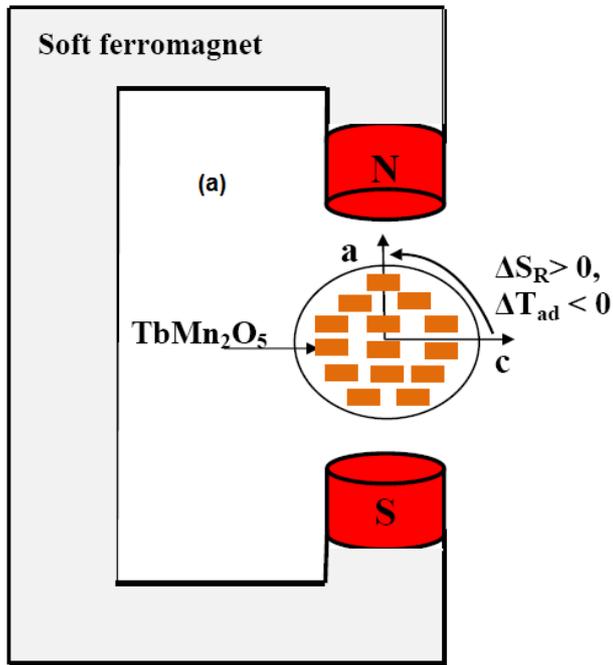

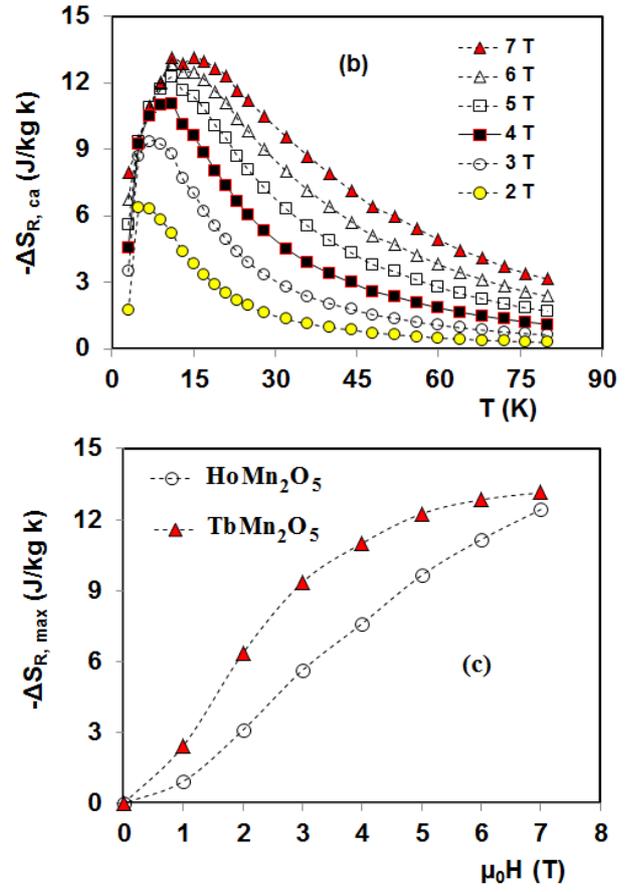

**Figure 7**

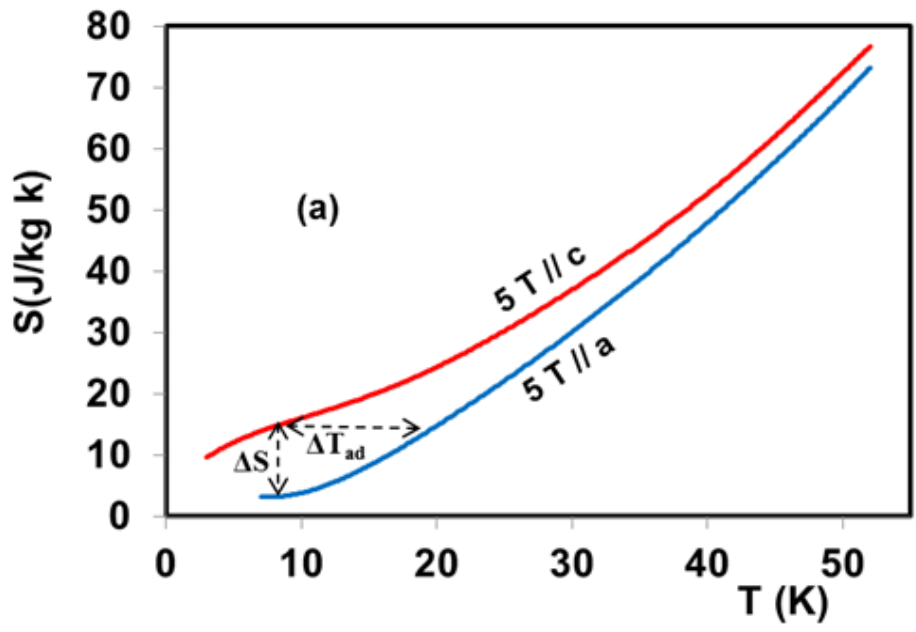

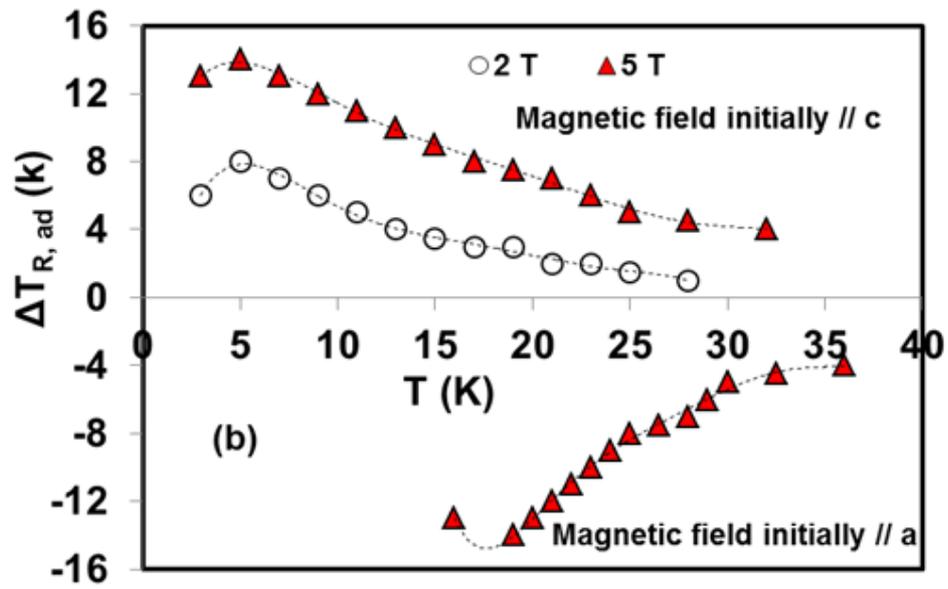

**Figure 8**

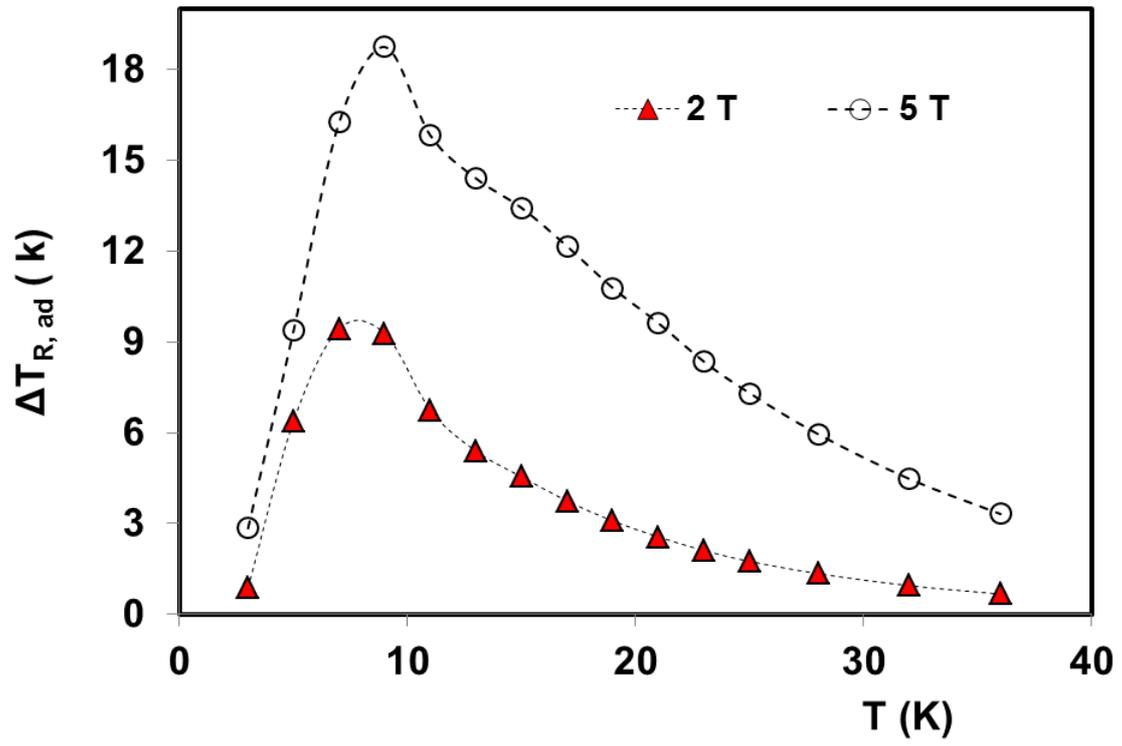

**Figure 9**

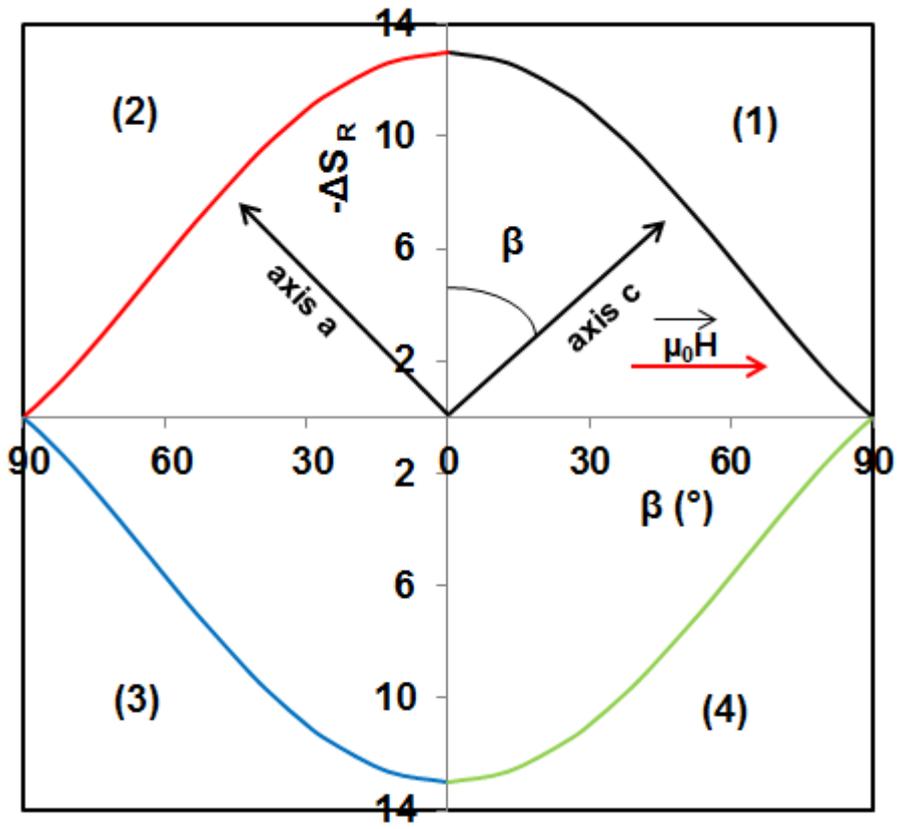

**Figure 10**